\documentclass{ametsoc}

\usepackage{longtable}
\usepackage{amsmath,empheq}
\usepackage{supertabular,booktabs}
\nolinenumbers
\journal{jas}

\bibpunct{(}{)}{;}{a}{}{,}

\title{Statistical theory on the analytical form of cloud particle size distributions}

\authors{Wei Wu\thanks{Current affliation: Department of Atmospheric Science, University of Wyoming, Laramie, WY.}}

\affiliation{Department of Atmospheric Sciences, University of Illinois at Urbana-Champaign, Urbana, IL \\ National Center for Atmospheric Research, Boulder, CO}

\extraauthor{Greg M. McFarquhar\correspondingauthor{Greg McFarquhar, Cooperative Institute for Mesoscale Meteorological Studies, University of Oklahoma, 120 David L. Boren Blvd, Norman, OK.}}

\extraaffil{Cooperative Institute for Mesoscale Meteorological Studies, Norman, OK \\ School of Meteorology, University of Oklahoma, Norman, OK}

\extraauthor{ }
\extraauthor{ }
\extraauthor{ }
\extraauthor{This work has been submitted for publication. Copyright in this work may be transferred without further notice, and this version may no longer be accessible.}

\email{mcfarq@ou.edu}

\abstract{Several analytical forms of cloud particle size distributions (PSDs) have been used in numerical modeling and remote sensing retrieval studies of clouds and precipitation, including exponential, gamma, lognormal, and Weibull distributions. However, there is no satisfying physical explanation as to why certain distribution forms preferentially occur instead of others. Theoretically, the analytical form of a PSD can be derived by directly solving the general dynamic equation, but no analytical solutions have been found yet. Instead of a process level approach, the use of the principle of maximum entropy (MaxEnt) for determining the analytical form of PSDs from the perspective of system is examined here. \\
\indent MaxEnt theory states that the probability density function with the largest information entropy among a group satisfying the given properties of the variable should be chosen. Here, the issue of variability under coordinate transformations that arises using the Gibbs/Shannon definition of entropy is identified, and the use of the concept of relative entropy to avoid these problems is discussed. Focusing on cloud physics, the four-parameter generalized gamma distribution is proposed as the analytical form of a PSD using the principle of maximum (relative) entropy with assumptions on power law relations between state variables, scale invariance and a further constraint on the expectation of one state variable (e.g. bulk water mass). The four parameter generalized gamma distribution is very flexible to accommodate various type of constraints that could be assumed for cloud PSDs.}

\begin{document}

\maketitle


\section{Introduction}
Various analytical forms of cloud particle size distributions (PSDs), such as exponential \citep{marshall1948distribution}, gamma \citep[e.g.,][]{borovikov1963cloud, ulbrich1983natural}, lognormal \citep[e.g.,][]{feingold1986lognormal,tian2010study} and Weibull distributions \citep[e.g.,][]{zhang1994simple, liu1995size}, have been used in numerical models and remote sensing retrieval algorithms. These functional forms of the distribution and the choice of free parameters characterizing the distribution have been typically determined on the basis of what provides the best match to in-situ observations. The scaling technique, as an alternative approach to describe cloud PSDs based on observational data, has been used recently to derive parameters characterizing a PSD by assuming a limited number of degrees of freedom and a ``universal distribution'' without stating its exact analytical form \citep[e.g.][]{testud2001concept,lee2004general}. Without considering the number of degrees of freedom needed to characterize a PSD, determining the analytical form of the ``universal distribution'' used in the scaling approach is a challenging question. Although many different analytical forms of cloud PSDs have been proposed, no study has yet provided an adequate physical explanation as to why a certain functional form is preferred over another. Therefore, the choice of functional form varies from study to study, complicating the comparison of PSD parameters derived from different field campaigns and from model parameterization schemes. It is not known if the choice of functional form should vary with environmental conditions.

A theoretical way to find an analytical form of a PSD is to solve the general dynamic equation describing the particle system, given by
\begin{multline}\label{eq:1}
\frac{\partial {n}(v,t)}{\partial t}
   =  - {n}(v,t) \int_0^{+\infty}  {K}(v, u)  {n}(u,t) \,du + \frac{1}{2}\int_0^{v}  {K}(u, v-u)  {n}(u,t)  {n}(v-u, t) \,du \\ 
       + \int_0^{+\infty}  {L}(v, u)  {n}(u,t) \,du - \frac{ {n}(v,t)}{v} \int_0^{v} u  {L}(v, u) \,du +  {SC}(v,t) -  {SK}(v,t) 
\end{multline} 
where ${n}(v,t)$ is the number distribution function for particles with volume $v$ at time $t$, ${K}(v,u)$ and ${L}(u, v)$ are the collection kernel and breakup kernel for particles with volumes $v$ and $u$, ${SC}(v,t)$ is the source term, and ${SK}(v,t)$ the sink term. All variables used in this paper are also defined in the Appendix. This form of the equation can be used for several different types of particles in a mixed particle system, such as ice particles with varying shapes and liquid particles. For the particle system of a single species (e.g. purely liquid clouds), one equation is sufficient. Unfortunately, Eq (1) can only be solved analytically for constant, additive or multiplicative kernels. Therefore, even for the simplest case of liquid clouds without nucleation, sedimentation and breakup, no analytic form for a cloud PSD has been found when a geometric collection kernel is used \citep{drake1972general}. When more complex processes acting in ice or mixed phase clouds are considered (e.g., sublimation, aggregation, melting, riming, deposition, etc.), the equation is even more difficult to solve and an analytic solution cannot be contemplated at this time. Because analytic solutions have not been possible, numerical methods have been used to determine PSDs in bin resolved models. However, the derived PSDs are very sensitive to even the representation of processes in liquid-phase clouds, such as the choice of raindrop breakup kernel \citep{srivastava1971size,srivastava1982simple,list1990role,hu1995evolution, mcfarquhar2004new}, with the collision-induced breakup parameterization determining the shape of modeled PSD. There are sensitivities to the representation of even more processes for ice or mixed-phase clouds.
 
A statistical theory is another viable way to determine the form of PSDs. Here the mass or size of every particle is considered as a random variable acting under the influence of stochastic processes from a statistical perspective, even though each individual particle follows physical laws. An example is the use of statistical mechanics in the field of the thermodynamics where every molecule follows physical laws, but the collections of molecules are described by statistical properties (e.g., the temperature represents the average kinectic energy of molecules). One promising statistical theory for determining cloud PSDs is the principle of maximum entropy \citep[MaxEnt,][]{jaynes1957informationa,jaynes1957informationb}. MaxEnt theory states that for a group of probability density functions (PDFs) that satisfy given properties of the variable, the PDF with largest information entropy for this variable should be chosen. Thus, a uniform distribution function (most uncertain) is selected if no other properties are specified. But, if the mean of the distribution is prescribed, the exponential distribution is the most probable distribution, following the same logic as used to derive the Maxwell-Boltzmann distribution in statistical mechanics. If both the mean and variance are prescribed, the most probable distribution is the normal distribution. The concept of MaxEnt has been used widely in physics \citep[e.g.,][]{rose1990statistical, antoniazzi2007maximum}, mechanical engineering \citep[e.g.,][]{sellens1985prediction, li1991comparison, berger1996maximum}, image processing \citep[e.g.,][]{wernecke1977maximum, skilling1984maximum}, machine learning \citep[e.g.,][]{rosenfeld1996maximum, berger1996maximum}, ecology \citep[e.g.,][]{phillips2004maximum,phillips2006maximum,banavar2010applications}, economics \citep[e.g.,][]{cozzolino1973maximum, buchen1996maximum}, and even in atmospheric sciences for representing cloud microphysics \citep[e.g.,][]{zhang1994simple, liu1995size, yano2016size} and turbulent flows \citep[e.g.,][]{majda2006nonlinear, craig2006fluctuations,verkley2009energy, verkley2011maximum, verkley2016maximum}. Its use in the study of spray PSDs in mechanical and material engineering is closely related to its use in the study of cloud PSDs. \cite{xianguo1987droplet}, \cite{dumouchel2006new}, and \cite{lecompte2008capability} employed MaxEnt to derive analytical forms of spray PSDs, and \cite{dechelette2011drop} has a comprehensive review on the application of MaxEnt to spray PSDs. Some applications of statistical mechanics may not state the principle of MaxEnt explicitly, but similar methods have been employed by \cite{griffith1943theory} to explain the particle size distribution in a comminuted system, and by \cite{lienhard1964statistical} to explain the unit hydrograph in hydrology. Thus, they are considered the same approach. 

The problem of determining PSDs in cloud physics is very similar to the problems in these other fields. For numerical models simulating clouds with bulk microphysics schemes, only a number of moments of the PSD are predicted. For example, many schemes prognose the mass and number concentration. Other moments of a PSD are then calculated using the assumed form of the PSD and assumptions about various constants describing these distribution forms \citep{thompson2004explicit, morrison2005new, seifert2006two, morrison2015parameterization}. These other moments include radar reflectivity and extinction. Thus, for developing parameterizations of cloud microphysics, there are some constraints on the properties of PSDs, exactly the type of scenario where MaxEnt can be used. Using MaxEnt, \cite{zhang1994simple} and \cite{liu1995size} introduced the Weibull distribution as the analytical form of PSDs assuming constraints on the surface area and mass, respectively. Their derived PSD forms differ on the parameters characterizing the Weibull distribution due to their different assumptions. \cite{yano2016size} extended the assumptions on the PSDs to include constraints on the mean maximum dimension and sedimentation flux of droplets, and examined the impact of these assumptions using idealized simulations, laboratory and observational datasets. All prior studies applying MaxEnt to cloud PSDs used the Gibbs/Shannon form of entropy.  However, the Gibbs/Shannon entropy is not invariant under coordinate transformation, and therefore different PSD forms can be derived using the same assumptions, as discussed in detail in Section 3. To solve these problems, different formalism of entropy is needed as \cite{jaynes1963information,jaynes1968prior} noted. 

This paper applies the form of (relative) entropy proposed by \cite{jaynes1963information,jaynes1968prior} to cloud PSDs. The problem of Gibbs/Shannon entropy is discussed in Section 3 after a brief review of MaxEnt in Section 2. Based on the form of (relative) entropy and several plausible assumptions about the cloud system, the four parameter generalized gamma distribution is proposed as the most reasonable analytical form of cloud PSDs in section 4. The properties of the generalized gamma distribution are summarized in section 5. The applications of the four-parameter generalized gamma distribution to in-situ observed liquid and ice clouds PSDs are investigated in section 6. The principle findings of the study and directions for future work are summarized in section 7.

\section{MaxEnt and its rationale for cloud physics}

MaxEnt theory was first proposed by \cite{jaynes1957informationa,jaynes1957informationb} to explain the classical Maxwell-Boltzmann distribution. The same principle has also been applied to Fermi-Dirac statistics and Bose-Einstein statistics and non-equilibrium statistical mechanics \citep{jaynes1963information, jaynes1968prior,dougherty1994foundations, banavar2010applications}. In statistical mechanics, if it is assumed that if there are $N_i$ particles in the $i$th energy state $E_i$, the total energy of the system $E$ is given by
\begin{equation}\label{eq:16}
E = \sum_{i=1}^{n} N_i E_i 
\end{equation} 
where there are $n$ total energy states with the total number of particles in the ensemble $N$ given by the summation of all particles in each energy state expressed by
\begin{equation}\label{eq:16}
N = \sum_{i=1}^{n} N_i .
\end{equation} 
The number of microscopic configurations in which the $N$ particles can be distributed over the $n$ different energy states, $W$, is given by
\begin{equation}\label{eq:16}
W=\frac{N!}{N_1 ! N_2 ! \cdots N_n !}.
\end{equation} 
Boltzmann defined the entropy as $S_B=k_B\ln(W)$, where $k_B$ is the Boltzmann constant \citep{pathria2011statistical}. Greater $W$ means a larger number of microscopic configurations of the $N$ particles distributed over the $n$ different energy state. $S_B$ monotonically increases with $W$ and is a measure of disorder: the greater the number of microscopic configurations in the system, the more uncertain the system can be. Using Sterling's formula
\begin{equation}\label{eq:16}
ln(n!) = nln(n) - n + O(ln(n)),
\end{equation} 
Boltzmann's entropy becomes
\begin{equation}\label{eq:16}
S_B= k_B ln(W) = k_B[  ln(N!) - \sum_{i=1}^{n} ln(N_i!) ]  \approx -k_BN\sum_{i=1}^{n} \frac{N_i}{N}ln(\frac{N_i}{N})   = -k_B N \sum_{i=1}^{n} p_i \ln(p_i) = N S, 
\end{equation} 
where $p_i = \frac{N_i}{N}$ is the probability of particles in the $i$th energy state, and $S=-k_B \sum_{i=1}^{n} p_i \ln(p_i)$ is Gibbs' form of entropy, which is the same form as Shannon's information entropy except for the inclusion of the Boltzmann constant \citep{shannon1948}. Assuming that there is a solution, denoted by $\bar{N}_i$ (or $\bar p_i$) that maximizes $W$ and therefore $S$, it can be proved using Eq (6) and the definition of Boltzmann entropy that
\begin{equation}\label{eq:16}
\frac{W_{max}}{W} = e^{\frac{N}{k_B}(S_{max}-S)}. 
\end{equation} 
Since $N$ is a very large number and $k_B$ is a very small number in the context of statistical mechanics, $W_{max}$ will be much larger than any other $W$ achieved with other $N_i$, indicating any other $p_i$ that deviates from $\bar p_i$ has significantly fewer microscopic configurations. For example for a mole of gas, there are $N_A$ (Avogadro constant, $6.02\times10^{23}$  mol$^{-1}$) particles, and the large ratio of $W_{max}$ to other $W$ rules out the possibility of other distributions. This is an important property of entropy.

The derivation of $\bar{N}_i$ or $\bar p_i$ is an optimization problem, $\operatorname*{arg\,max}_{N_i} \ln(W)$ subject to Eq (2) and Eq (3), that can be derived using the method of Lagrange multipliers, where
\begin{equation}\label{eq:13}
d \ln(W) - \lambda_0 (\sum_{i=1}^{n} N_i  - N ) - \lambda_1 (\sum_{i=1}^{n} N_i E_i  - E ) = 0
\end{equation} 
where $\lambda_0$ and $\lambda_1$ are the Lagrange multipliers. By using Stirling's approximation, Eq (8) becomes
\begin{equation}\label{eq:13}
\begin{aligned}
    & \sum_{i=1}^{n} - \ln(N_i) d N_i - \lambda_0 (\sum_{i=1}^{n} d N_i) - \lambda_1 (\sum_{i=1}^{n} dN_i E_i) = \sum_{i=1}^{n} ( - \ln(N_i) - \lambda_0  - \lambda_1 E_i ) d N_i  = 0 
\end{aligned}
\end{equation} 
so that
\begin{equation}\label{eq:16}
\bar p_i = \frac{\bar N_i}{N} = C e^{-\lambda_1 E_i} , \text{where } C = C_0 e^{-\lambda_0},
\end{equation} 
which is the Maxwell-Boltzmann distribution \citep{pathria2011statistical}. The Lagrange multipliers $\lambda_0$ and $\lambda_1$ can be obtained by substituting $N_i$ in Eq (2) and Eq (3).  

Based on the above arguments, \cite{jaynes1957informationa,jaynes1957informationb} argued that for a group of PDFs that satisfy the given properties of a variable $x$, the PDF with largest information entropy \citep{shannon1948} should characterize the variable, with statistical mechanics being just one example of this principle applied to an ideal gas. The methodology can be generalized using a continuous distribution to characterize the variable
\begin{equation}\label{eq:16}
\begin{aligned}
 & \operatorname*{arg\,max}_{P(x)} - \int_{0}^\infty {P(x) lnP(x) dx} \\ 
 & \text{subject to ($nc+1$) constraints: } \int_0^\infty f_k(x) P(x) dx = F_k
 \end{aligned}
\end{equation} 
where $P(x)$ is the probability that state variable $x$ will occur, and the $nc+1$ constraints are expressed in the form of fixed expectation of $f_k(x)$ with $k$ = 0, 1, 2..., $nc$. For $k$=0, $f_0(x)$=1 and $F_0$=1 are chosen as the normalization condition for the PDF. Since the 0th constraint is valid for every PDF, only $nc$ other constraints need to be given explicitly. Therefore, the number of given constraints is denoted as $nc$. The value of $nc$ is determined by the knowledge of the system that is being considered, and can vary according to the behavior of the particular system that is being modeled. To get the maximum of $S(x) = - \int_{0}^\infty {P(x) lnP(x) dx}$ with these constraints, the method of Lagrange multipliers can be applied as before with the discrete sums so that the Lagrange function $L(x, \lambda_1,  \lambda_2, ...,  \lambda_k)$ is expressed by
\begin{equation}\label{eq:13}
L(x, \lambda_1,  \lambda_2, ...,  \lambda_k) \equiv -\int_0^\infty P(x)lnP(x) - \sum_{k=0}^{nc} \lambda_k (\int_0^\infty f_k(x) P(x) dx - F_k )
\end{equation} 
where $k$ = 0, 1, 2, ..., $nc$. Then the general result can be solved so that
\begin{equation}\label{eq:13}
P(x) = \frac{1}{Z(\lambda_1, \lambda_2, .., \lambda_m)} exp(-\sum_{k=0}^{nc} \lambda_k f_k(x))
\end{equation} 
where the partition function $Z(\lambda_1, \lambda_2, .., \lambda_n) = \int exp(-\sum_{k=0}^{nc} \lambda_k f_k(x)) dx$.

Following this technique, the exponential distribution can be derived as the maximum entropy distribution if only the mean of the variable is known. The Weibull distribution can be further derived as the maximum entropy distribution if the mean of the power function of the variable is known. If both the mean and variance of a variable are known, the normal distribution will be the maximum entropy distribution. Similarly, the lognormal distribution will be derived if the mean and variance of the logarithm of the variable are known. \cite{kapur1989maximum} describes commonly used PDFs and their constraints.  

Statistical mechanics can be used to define the properties of a cloud just as it is used to define the properties of an ideal gas. Just as in thermodynamics where there are variables describing the microscopic and macroscopic state of the ideal gas, there are variables describing the microscopic and macroscopic properties of clouds as discussed previously, and the use of random variables is convenient so that statistical mechnics can be applied. The macroscopic states are mainly defined by the total number of cloud particles, the cumulative extinction (projected area) of all cloud particles, the bulk liquid or ice water content of all particles in a distribution and other bulk microphysical properties. The microscopic states are described by the size, area and mass of the individual hydrometeors. This is analogous to the case of an ideal gas, where even though there are numerous realizations of velocity for each individual gas molecule, the Maxwell-Boltzmann distribution, which has the largest entropy, has the largest number of microscopic configurations of molecule speeds and hence characterizes the distribution. Similarly, for clouds, the PDF with the maximum entropy also has the largest number of microscopic configurations of cloud particles distributed over different sizes. A key question in the application of statistical mechanics to distributions of cloud particles is how many particles are needed to make the method robust, because there are inevitably fewer cloud particles than gas molecules. If it is assumed that the total number concentration is $N_t$, then the total number of cloud particles in a sample volume $V$ is $N=N_t V$. The volume should be sufficiently large to make $N$ large, but at the same time, not so large to exceed the typical volume of a cloud or a scale where there is a lot of horizontal or vertical inhomogeneity. Here a unit cloud volume ($V$) of 100m x 100m x 10m = $10^5 m^3$ is proposed as large enough. Assuming a concentration of $N_t\approx 100 cm^{-3}$, then $N=N_tV=10^{13}$ should be big enough to make the derivation solid. This volume is also small enough compared to typical model grid volume or radar sample volumes. 

In cloud physics, the number distribution function is expressed as $N(D)$, which can be normalized by $N_t=\int_0^\infty N(D) dD$ to define the number distribution probability density function expressed by 
\begin{equation}\label{eq:13}
P(D)=\frac{N(D)}{N_t}. 
\end{equation} 
Thus, the MaxEnt approach can be applied in the study of cloud PSDs, and its use in cloud physics has been discussed by \cite{zhang1994simple}, \cite{liu1995size} and \cite{yano2016size}. However, there are problems directly applying MaxEnt to cloud PSDs as discussed in section 3. Previous studies chose the particle maximum dimension ($D$) or particle mass ($m$) as the state variable $x$, with all assuming two constraints: 1) the constraint of total particle number concentration;  and 2) a constraint of mean maximum dimension \citep{yano2016size}, total surface area \citep{zhang1994simple}, total bulk water content \citep{liu1995size}, or mass flux \citep{yano2016size}. The constraints apply to bulk properties, which are integrations of particle properties over size. The derived PSD forms maximizing the entropy are then special cases of Eq (13), with $nc=1$, expressed by
\begin{equation}\label{eq:13}
P(x) = \frac{1}{Z(\lambda)} exp(-\lambda_1 f_1(x))
\end{equation} 
where $x$ could be $D$ or $m$, and $f_1(x)$ could be $D$, $A$, $m$, or $m v$ ($v$ is the fall speed of a particle). The $f_1(x)$ is typically written as a power law function of $x$. An example for $f_1(x)$ is $\alpha D^\beta$, where $x$ is $D$.  Note that the PDF as a function of $D$ and the PDF as a function of any other variable $x$ can be converted, so that the number distribution function can be expressed
\begin{equation}\label{eq:13}
N(D) = N_t P(D) = N_t P(x)\frac{dx}{dD} =   \frac{N_t}{Z(\lambda)} exp(-\lambda_1 f_1(x))\frac{dx}{dD}.
\end{equation} 
Usually, any other state variable $x$ and the particle maximum dimension $D$ are assumed to be related through a power law (e.g., $x=aD^b$), so that Eq (16) can be rewritten as  
\begin{equation}\label{eq:13}
N(D) =  \frac{N_t ab}{Z(\lambda)} D^{b-1}exp(-\lambda_1 f_1(aD^b)),
\end{equation} 
which is a special case of the generalized gamma distribution, with parameter $b$ being the power parameter in mass-dimensional relations. For droplet size distributions, the parameter $b$ is 3.

\section{Problems using Gibbs/Shannon entropy and the concept of relative entropy}
Eq (17) is a general solution for the functional form of cloud PSDs maximizing the entropy content as long as one constraint is given explicitly. However, it can be proven that a different PSD form can be derived using the same constraint when using the Gibbs/Shannon entropy. For example, below it is shown that the same constraints used in \cite{liu1995size} can be employed to derive a different PSD than the one derived in \cite{liu1995size}. It should be noted that all the forms derived in \cite{zhang1994simple} and \cite{yano2016size} suffer the same problem. Assuming that the total bulk number concentration $N_t$ and total bulk water mass content $TWC$ are constraints and using mass $m$ as the variable characterizing particles, \cite{liu1995size} showed that the MaxEnt distribution was given by
\begin{equation}\label{eq:16}
N(m)= C_{1} exp(-\lambda_{1} m) 
\end{equation} 
where $C_1=\frac{N_t^2}{TWC}$ and $\lambda_1=\frac{N_t}{TWC}$ are the distribution parameters. This distribution can be rewritten in term of $D$ using an assumed mass-dimensional relation $m=\alpha D^{\beta}$ as
\begin{equation}\label{eq:16}
N(D)= \bar C_{1} D^{\beta-1} exp(-\bar \lambda_{1} D^{\beta}) 
\end{equation} 
where $\bar C_1=\frac{\alpha \beta N_t^2}{TWC}$ and $\bar \lambda_1=\frac{\alpha N_t}{TWC}$ are the distribution parameters.

However, if the PDF is characterized in terms of $D$ instead, and the same two constraints are applied as expressed by 
\begin{equation}\label{eq:20}
\int_0^\infty N(D) dD =\int_0^\infty N_t P(D) dD =N_{t} 
\end{equation} 

\begin{equation}\label{eq:21}
\int_0^\infty \alpha D^{\beta} N(D) dD = \int_0^\infty \alpha D^{\beta} N_t P(D) dD =TWC
\end{equation} 
the MaxEnt distribution becomes
\begin{equation}\label{eq:22}
N(D)= \tilde C_{1} exp(-\tilde \lambda_{1}D^{\beta}) 
\end{equation} 
where $\tilde C_{1}=\frac{N_t \beta (\frac{\alpha N_t}{\beta TWC})^{\frac{1}{\beta}}}{\Gamma(\frac{1}{\beta})}$ and $\tilde \lambda_1=\frac{\alpha N_t}{\beta TWC}$ are the distribution parameters. By comparing Eq (19) and Eq (22), it is found that two different analytical forms of PSDs are derived using the same assumption. In fact, a different analytical form of the PSD can be derived whenever the state variable $x$ characterizing the cloud particles changes. This is due to the fact that the Gibbs/Shannon entropy is not invariant under the transformation of variables \citep{jaynes1963information,jaynes1968prior}. Thus, \cite{jaynes1963information,jaynes1968prior} proposed another definition of entropy, typically called relative entropy, that makes entropy invariant under variable transformations. The relative entropy is mathematically sound and physically meaningful as discussed below.

The definition of relative entropy proposed by \cite{jaynes1963information, jaynes1968prior}, $S_r$, is expressed by
\begin{equation}\label{eq:11}
S_r(x)= - \int_{0}^\infty {P(x) ln\frac{P(x)}{I(x)} dx},
\end{equation} 
where $I(x)$ is called the invariant measure, or a prior distribution that represents an initial guess of what the distribution should be. This relative entropy $S_r(x)$ has also been called Kullback-Leibler divergence, which is a measure how a PDF diverges from a prior distribution. When $I(x)$ is a uniform distribution, the definition of relative entropy is identical to the Gibbs/Shannon entropy minus a constant. For systems where coordinate transforms are important, the uniform distribution is not a good prior distribution. Therefore, a form that is invariant under coordinate transforms needs to be used for the generalized development. It can be shown that $S_r$ is invariant under coordinate transformation ($x \rightarrow y$, where $y=g(x)$), because
\begin{equation}\label{eq:16}
S_r(y)= - \int_{0}^\infty {P'(y) ln\frac{P'(y)}{I'(y)} dy} = - \int_{0}^\infty {P(x) ln\frac{P(x)}{I(x)} dx} = S_r(x),
\end{equation} 
with $P'(y)=P(x)\frac{dx}{dy}$ and $I'(y)=I(x)\frac{dx}{dy}$. 

To maximize $S_r(x)$ with given constraints, the method of Lagrange multipliers is again used so that
\begin{equation}\label{eq:13}
L \equiv -\int_0^\infty P(x)ln\frac{P(x)}{I(x)} - \sum_{k=1}^{nc} \lambda_k (\int_0^\infty f_k(x) P(x) dx - F_k )
\end{equation} 
where $k$ = 0, 1, 2, ..., $nc$ and the maximum (relative) entropy distribution is solved in the form
\begin{equation}\label{eq:13}
P(x) = \frac{1}{Z(\lambda_1, \lambda_2, .., \lambda_{nc})} I(x) exp(-\sum_{k=0}^{nc} \lambda_k f_k(x))
\end{equation} 
where the new partition function is $Z(\lambda_1, \lambda_2, .., \lambda_{nc}) = \int_0^\infty I(x) exp(-\sum_{k=0}^{nc} \lambda_k f_k(x)) dx$.

\section{Application to cloud PSDs}
The relative entropy is invariant under coordinate transformations, and the distribution derived maximizing this definition of entropy is consistent with the same constraint regardless of the variable used to characterize the PDF. However, before the theory can be applied to any system, the appropriate constraints and invariant measure $I(x)$ are needed. These can only be obtained from an understanding of the system studied. To apply the theory to cloud physics, the first step is to determine the constraints for a cloud. \cite{yano2016size} used observed and simulated datasets to evaluate constraints of mean maximum dimension, bulk extinction, bulk water content, and bulk mass flux. This paper does not examine the use of different constraints as did \cite{yano2016size}, but instead focuses on the general application of the new definition of entropy. Unlike the Gibbs/Shannon entropy used in previous studies, the choice of state variable $x$ is not important for $S_r$, as the invariant measure $I(x)$ will adjust accordingly. In this study, the particle maximum dimension ($D$) is chosen as the state variable $x$ of the cloud. The number distribution function $N(D)$, following Eq (26), can thus be expressed by
\begin{equation}\label{eq:27}
N(D) = \frac{N_t}{Z(\lambda_1, \lambda_2, .., \lambda_{nc})} I(D) exp(-\sum_{k=0}^{nc} \lambda_k f_k(D))
\end{equation} 
where the number of constraints is usually equal or larger than 1. Eq (27) is the general form of $N(D)$, and the number of constraints as well as the corresponding constraint function $f_k(D)$ need to be assumed to derive a specific form for $N(D)$. For this study, one constraint is assumed. In future studies, more than one constraint can be used if the solution using one constraint is not well validated against observations. If it is assumed that this one constraint and the constraint function is represented as a power law with particle maximum dimension ($f_1(D) = aD^b$) following \cite{zhang1994simple}, \cite{liu1995size}, and \cite{yano2016size}, Eq (27) then becomes
\begin{equation}\label{eq:28}
N(D) = \frac{N_t}{Z(\lambda_1)} I(D) e^{-\lambda_1 a D^b} = \frac{N_t}{Z(\lambda_1)} I(D) e^{-\lambda D^b}
\end{equation} 
where $\lambda = \lambda_1a$.

The next step applying MaxEnt theory is to determine the invariant measure $I(D)$, which must be provided from a knowledge of the underlying physics. \cite{jaynes1968prior} provided guidelines to choose the invariant measure based on the transformation group, and \cite{jaynes1973well} showed an example using the transformation group. The basic idea is that the shape of the invariant measure should be invariant in two different systems. In particular for this case, the shape of the invariant measure should not change with the volume of cloud studied. Assume two volumes of the same clouds: cloud A with total volume $V_A$ and cloud B, a subset of the cloud A, with total volume $V_B=\kappa_1 V_A$, $\kappa_0<\kappa_1\le 1$. Hereafter, the properties of volume A and volume B are denoted with the subscripts A and B respectively. Here, $\kappa_1$ can not be too small since a large number of particles is needed for the application of statistical mechanics; hence $\kappa_0$ is used for a lower bound instead of 0. Here, $\kappa_0=0.1$ should be large enough, which will give the number of cloud particles in the volume B around $10^{12}$ if volume A is as discussed in section 2. 

For volume $V_A$, the total mass is $TWC\times V_A$. Therefore no particles larger than $D_{maxA}$ are possible, where $\alpha {D_{maxA}}^\beta=TWC\times V_A$ with $\alpha$ and $\beta$ the $m-D$ relation parameters with $m=\alpha D^{\beta}$ the mass of an individual cloud particle. Thus, $I_{A}(D) =0$ for $D>D_{maxA}$. For cloud A, the prior probability is $I_{A}(D)$ with $\int_0^{D_{maxA}} I_{A}(D) dD = 1$. For cloud B, the volume will be $V_B = \kappa_1 V_A$, the maximum particle size will be $D_{maxB} = \kappa D_{maxA}$ ($\kappa = \kappa_1^{1/\beta}$), and the prior probability $I_{B}(D)$ satisfies $\int_0^{D_{maxB}} I_{B}(D) dD = 1$. A new scaled dimensionless variable $x=\frac{D}{D_{maxA}}$ is defined to scale $I_A(D)$ into the range [0,1] so that 
\begin{equation}\label{eq:13}
\int_0^{D_{maxA}} I_{A}(D) dD = \int_0^{1} I_{A}(x D_{maxA}) d(x D_{maxA}) = \int_0^{1} D_{maxA} I_{A}(x D_{maxA}) dx = \int_0^{1} f_A(x) dx = 1
\end{equation} 
where $f_A(x)=D_{max} I_{A}(x D_{maxA})$, and similarly $f_B(y)=D_{maxB} I_{B}(y D_{maxB})$ where $y=\frac{D}{D_{maxB}}$. Because of scale invariance, the scaled PDFs $f_A(x)$ and $f_B(y)$ over the same range of [0, 1] should be the same, meaning that 
\begin{equation}\label{eq:13}
\begin{split}
&        f_A(x) = f_B(x)\\
& \text{or } D_{maxA} I_{A}(x D_{maxA}) = D_{maxB} I_{B}(x D_{maxB})  \\
& \text{or } I_{A}(D) = \kappa I_{B}(\kappa D).
\end{split}
\end{equation} 
This is the scale invariance that the cloud system must satisfy in order for two different volumes to have the same shape of invariant measure. Following the formula for conditional probability, for any $D$ that is within the range of (0, $\kappa D{maxA}$], it can be shown that   
\begin{equation}\label{eq:13}
I_{A}(D) = I_{B}(D) \int_0^{\kappa D_{maxA}} I_{A}(u) du. 
\end{equation} 
Eq (31) is the standard conditional probability formula, and will hold whether or not any transformation invariance is assumed. 


Combining the invariance requirement Eq (30) and the conditional probability relation Eq (31), it is determined that
\begin{equation}\label{eq:13}
\kappa I_{A}(\kappa D) = I_{A}(D) \int_0^{\kappa D_{maxA}} I_{A}(u) du.
\end{equation} 
Differentiating Eq (32) with respect to $\kappa$ gives another form of Eq (32), 
\begin{equation}\label{eq:13}
I_{A}(\kappa D) + \kappa \frac{\partial I_{A}(\kappa D)}{\partial D} D = I_{A}(D) I_{A}(\kappa D_{maxA})D_{maxA}.
\end{equation} 
This equation is still hard to solve since it involves both cloud A and cloud B, and therefore one parameter $\kappa$. Remember that cloud B is a subset of cloud A. By setting $\kappa=1$ to make cloud A and cloud B the same, the equation for one cloud (cloud A is chosen here, but it is the same to choose cloud B to solve first) yields
\begin{equation}\label{eq:13}
I_{A}(D) + \frac{\partial I_{A}(D)}{\partial D} D = I_{A}(D) I_{A}(D_{maxA})D_{maxA}  \rightarrow \frac{\partial I_A(D)}{\partial D}D = (I_A(D_{maxA})D_{maxA} -1 )I_A(D).
\end{equation} 
Solving the differential equation Eq (34), it can be shown that the most general solution is
\begin{equation}\label{eq:13}
I_A(D) = \frac{\mu+1}{{D_{maxA}}^{\mu+1}}D^\mu
\end{equation} 
where $\mu$, defined as $I_A(D_{maxA})D_{maxA}-1$, is a constant in the range of $-1< \mu < \infty$. The constant $\mu$ cannot be further determined by scale invariance. Using Eq (30), the invariant measure of cloud B is 
\begin{equation}\label{eq:13}
I_B(D) = \frac{\mu+1}{{(\kappa D_{maxA})}^{\mu+1}}D^\mu = \frac{\mu+1}{{D_{maxB}}^{\mu+1}}D^\mu
\end{equation} 

The form of invariant measure provided by Eq (35) (or Eq (36)) satisfies translational, rotational and scale transformations, typical transformations between coordinate systems as suggested by \cite{jaynes1968prior}. In this case, PSDs are described with a single dimension in a space related to particle maximum dimension, so no rotational transformation exists. Since no spatial variables are involved, the PDF does not change when the coordinate system is translated. Scale transformation is the last transformation to be satisfied. For two coordinate systems R and S, the length relates by $\bar \kappa$ with $D_R = \bar \kappa D_S$, the invariant measure for R is $I_R(D_R)$ with $\int_0^{D_{maxR}} I_R(D_R) dD_R= 1$ and the invariant measure for S is $I_S(D_S)$ with $\int_0^{D_{maxS}} I_S(D_S) dD_S = 1$. Since it is the same cloud observed, the relation $I_R(D_R) dD_R=I_S(D_S) dD_S$ holds, which means $\bar \kappa I_R(\bar \kappa D_S) = I_S(D_S)$. Eq (35) clearly satisfies this relation. So far the invariant measure provided by Eq (35) satisfies all the Abelian group transformations proposed by \cite{jaynes1968prior}.  

If Eq (35) is assumed to represent the invariant measure, combined with Eq (28), the final $N(D)$ is the four-parameter generalized (or modified) gamma distribution, given by
\begin{equation}\label{eq:a2}
N(D) = N_0 D^{\mu} e^{-\lambda D^b}
\end{equation}
where $N_0 = \frac{N_t C}{Z(\lambda_1)}$.

It should be noted that the derived PSD forms from \cite{zhang1994simple}, \cite{liu1995size} and \cite{yano2016size} are all special cases of Eq (37), so that this study is consistent with, but more general than, previous studies. It is also clear now why the two approaches to derive the PSD form in section 3 generate different results. Eq (19) and Eq (22) differ by $D^\mu$, which is the invariant measure. The first approach assumed a uniform invariant measure over particle mass and the second assumed a uniform invariant measure over particle size, and $\frac{dm}{dD}=\alpha \beta D^{\beta-1}$ is the difference.   

\section{Properties of generalized Gamma distribution}
The properties of the generalized gamma distribution are summarized in this section. The generalized (or modified) gamma distribution is a general form of a PSD, which can be simplified to an exponential, gamma or Weibull distribution in special cases. To the authors' knowledge, the generalized gamma distribution was first proposed by \cite{amoroso1925ricerche} to study income distribution, and later independently proposed by \cite{nukiyama1939experiment} for fitting the size distribution of sprays particles in mechanical and material engineering. \cite{stacy1962generalization} studied the mathematical properties of the generalized gamma distribution, and the properties relevant to cloud PSDs are summarized here. 

The cumulative distribution function for the generalized (modified) gamma distribution in the form of Eq (37) is
\begin{equation}\label{eq:a2}
F(D;N_0,\mu,\lambda,b) = \frac{N_0}{b\lambda\Gamma(\frac{\mu+1}{b})} \gamma(\frac{\mu+1}{b}, \lambda D^b)
\end{equation}
where $\gamma(s,x) = \int_0^x t^{s-1} e^{-t} dt$ is the lower incomplete gamma function. The $n$th moment can be calculated as
\begin{equation}\label{eq:a2}
M_n = E(x^n)=\frac{N_0}{b\lambda^{(\mu+1+n)/b+1}} \Gamma(\frac{\mu+1+n}{b}).
\end{equation}

For any variable $x$ that is related to $D$ through a power law (e.g., $x=c D^d$), it can also be represented by a generalized gamma distribution with the form
\begin{equation}\label{eq:a2}
N(x) = \frac{N_0}{c^{(\mu+1)/d}d} x^{\frac{\mu+1}{d}-1} e^{-\frac{\lambda}{c^{b/d}} x^{\frac{b}{d}}} 
\end{equation}

One main benefit of the four-parameter generalized gamma distribution is that it is invariant under coordinate transformations when characterizing a PSD. The same form applies to all power law variables, such as particle maximum dimension, area and mass. The lognormal distribution also has this property and this is one of the reasons \cite{feingold1986lognormal} recommended the lognormal distribution for PSDs. This property is not shared by the exponential distribution, gamma distribution or Weibull distribution. For example, \cite{seifert2006two} assumed the commonly used three-parameter gamma distribution over mass, which will turn into a four-parameter generalized gamma distribution over particle size. A second benefit is that the generalized gamma distribution can also simplify to a gamma distribution, Weibull distribution or even exponential distribution under certain circumstances. Third, the physical meaning of distribution parameters is more clear than parameters used in some empirical distribution functions used in previous studies. Due to the properties mentioned above, \cite{maur2001statistical} and \cite{petty2011modified} also proposed the use of generalized (or modified) gamma distribution without stating the underlying physical basis. 

\section{Testing with in-situ observed liquid and ice PSDs}
In this section, in-situ observed PSDs are fit to different analytical forms, including the gamma, Weibull, lognormal and generalized gamma distribution. The fitting in this section is used to test the application of four-parameter generalized gamma distribution in real clouds. 

An in-situ dataset collected by a two-dimensional cloud probe (2DC) and high sample volume spectrometer (HVPS) during the Midlatitude Continental Convective Clouds Experiment \citep[MC3E, ][]{JensenMC3E} are used for the fitting. \cite{wu2016impacts} describe how the data were collected and how the binary data were processed to generate cloud PSDs. Two different distributions were used in the analysis: a one-minute time period in liquid clouds and another one-minute period in ice clouds. The particle images were all manually checked to make sure no mixed-phase time periods existed in these two time periods. Liquid PSDs measured between 13:20:00-13:20:59 at a temperature of around 4 $^o$C are averaged, and the best fits to the different analytical functions listed in the legend of Fig. 1 were performed. Following \cite{mcfarquhar2015characterization}, the fitting technique minimized the $\chi^2$ differece between the fit and observed moments of $N(D)$ defined by 
\begin{equation}\label{eq:a2}
\chi^2 = \sum_{i=1}^{nm} [\frac{M_{fit,i}-M_{obs,i}}{\sqrt{M_{fit,i}M_{obs,i}}}]^2 
\end{equation}
where $M_{obs,i}$ is the $i$th moment of the observed PSD, and $M_{fit,i}$ is the $i$th moment of the fit PSD calculated using the assumed PSD form. Here the 0th, 3rd and 6th moments corresponding to total number concentration, bulk liquid water content and radar reflectivity were used in the fitting procedure to determine the parameters describing the gamma, Weibull and lognormal distributions. To determine the parameters of the generalized gamma distribution, the first moment, representing the mean particle size, was also used because four moments are required to describe the four-parameter of the generalized gamma distribution. All fit functions had $\chi^2 \leq 0.001$ in Eq (41), showing all fits provide good agreement between fit and measured moments. Further, the fit gamma, Weibull and generalized gamma distributions all appear similar to the observed PSD, while the lognormal fit seems to deviate further from the observed PSDs. The fit generalized gamma distribution has a $b$ parameter very close to 1 (0.99), so the fit curve is very close to the gamma distribution. This implies that the mean maximum dimension is the constraint for the liquid clouds in this time period.

\begin{figure*}[t]
\begin{center}
  \noindent\includegraphics[trim=2cm 0cm 3cm 0cm, clip, width=\linewidth,angle=0]{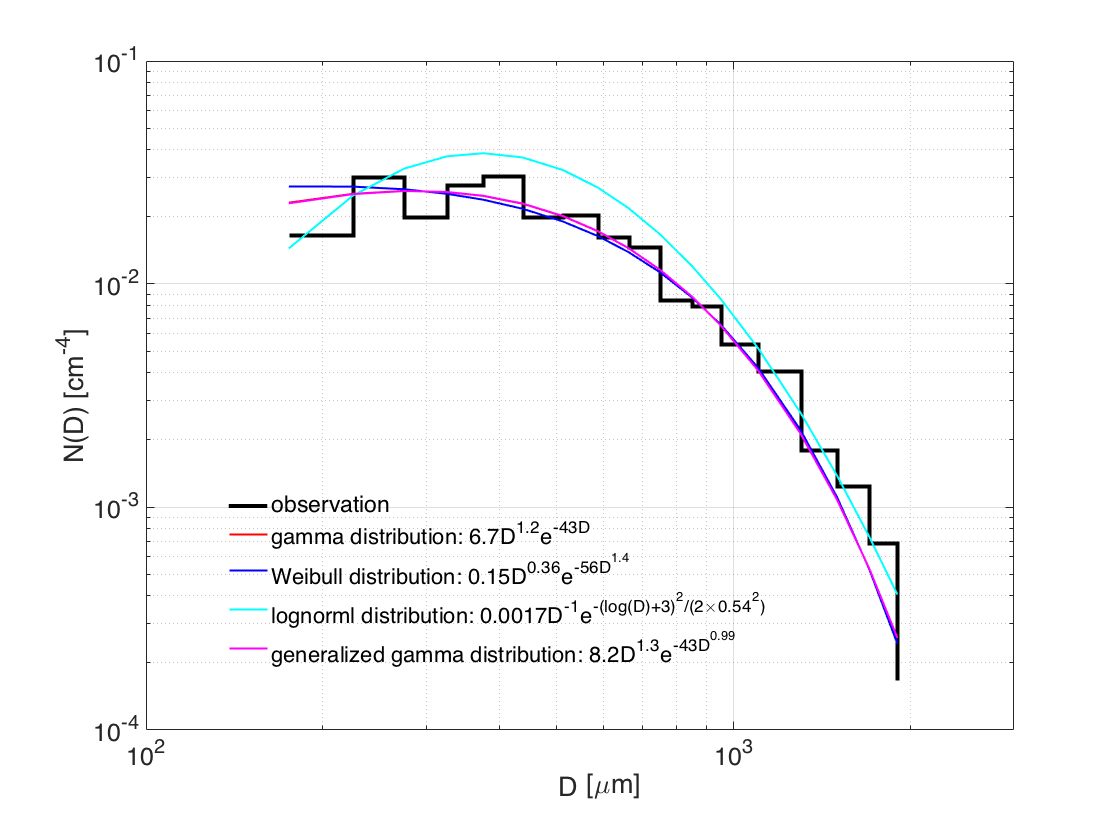}\\
  \caption{Sample in-situ liquid PSD $N(D)$ as function of $D$ (black) and fitted for gamma distribution (red), Weibull distribution (blue), lognormal distribution (cyan) and generalized gamma distribution (purple). The red curve is right under the purple curve. The fitted parameters are listed in the legend. }\label{f1}
\end{center}
\end{figure*}

Fits to the PSD measured in ice clouds from 15:55:00-15:55:59 at a temperature of around -10 $^o$C from the same flight were also conducted with the 0th, 2nd and 4th moments used to determine the fit parameters. For ice clouds, these approximately correspond to the total number concentration, bulk ice water content and radar reflectivity, respectively. Similarly, an additional moment, the 1st moment, was used to find the generalized gamma distribution fit parameters. Figure 2 shows the results of the fits that were performed. The $b$ parameter in the generalized gamma distribution is 0.39, and the fit curve is closer to the observed PSDs compared to gamma distribution and Weibull distribution.

\begin{figure*}[t]
\begin{center}
  \noindent\includegraphics[trim=1cm 0cm 3cm 0cm, clip, width=\linewidth,angle=0]{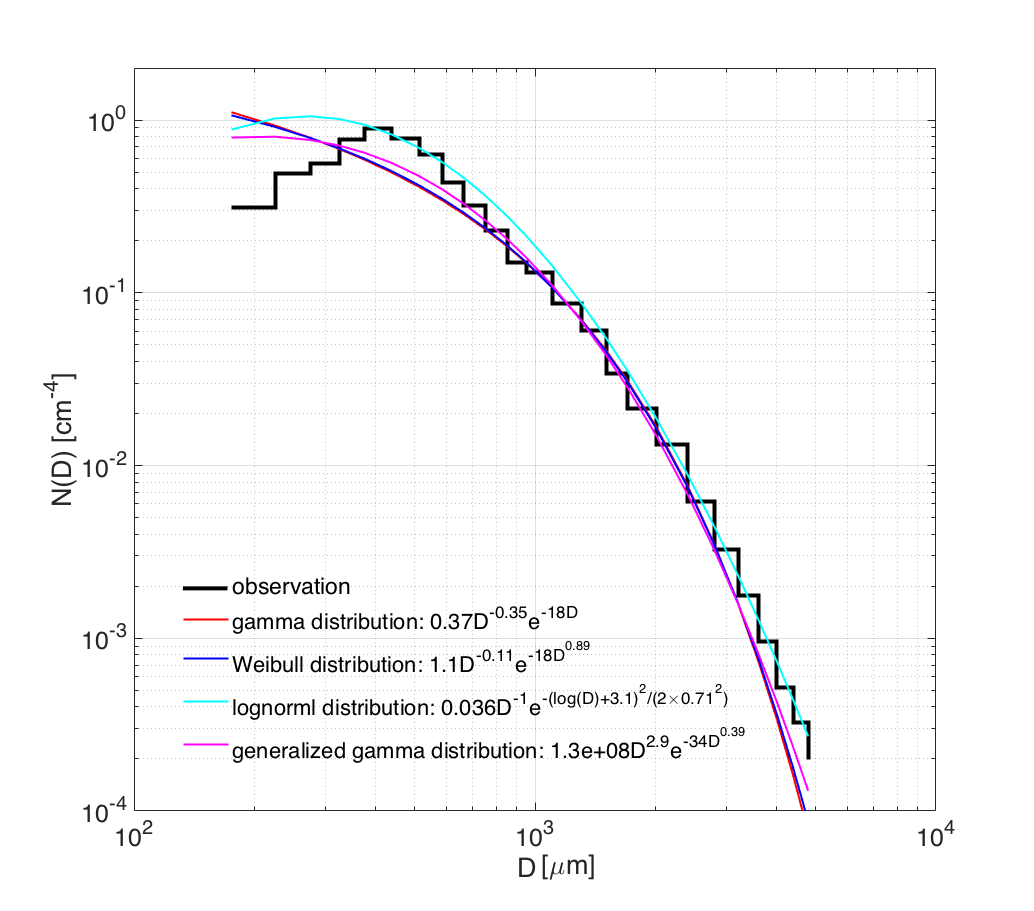}\\
  \caption{Same as Fig. 1, but for ice PSDs. }\label{f2}
\end{center}
\end{figure*}

\section{Conclusions and discussions}
Several analytical forms of cloud PSDs, such as exponential and gamma distribution functions, have been assumed in numerical models and remote sensing retrievals in past studies. However, no satisfying physical basis has yet been provided for why any of these functions characterize PSDs. The use of the principle of maximum entropy (MaxEnt) to find analytical forms of PSDs was examined here, building upon its use in prior studies \citep{zhang1994simple,liu1995size,yano2016size}. The main findings of this study are summarized as follows:

1). The definition of relative entropy, $S_r = - \int_{0}^\infty {P(x) ln\frac{P(x)}{I(x)} dx}$ which is invariant under coordinate transformations, was used to resolve an inconsistency in previous studies. The previous use of Gibbs/Shannon entropy allowed different PSD to be derived using the same constraint by simply using a different state variable $x$.  

2). The definition of relative entropy used in this study to determine a physical basis for a cloud PSD requires an assumption about an invariant measure $I(D)$, which is obtained from a physical understanding of the system studied. Here, it was shown that $I(D)$ can be obtained if invariance regarding group transformation is assumed.  

3). Assuming that the microscopic state variables that characterize the properties of cloud particles (e.g., particle maximum dimension, area, mass, fall speed) are related to each other through power laws, it was shown that if one constraint related to any state variable was assumed, a four parameter generalized gamma distribution can be derived. The state variable that needs to be used as a constraint is not yet well determined.

4). It was shown that if one state variable follows the generalized gamma distribution, all state variables having power law relations with the state variable must also follow the generalized gamma distribution.

5). Directly fitting in-situ observed PSDs using data obtained from optical array probes (OAPs) generates reasonable fits to the observed PSDs for all the analytical forms of PSD, even though the fit of generalized gamma distribution is slightly better. Due to the discrete nature of observed PSDs and large uncertainties for OAPs, parameters derived by directly fitting have large uncertainties.

Although the MaxEnt approach provides a physical basis for the form of the generalized four-parameter gamma distribution, it does not determine the values of parameters ($N_0$, $\mu$, $\lambda$ and $b$). These can only be determined using observational datasets. Among the four parameters, $b$ is particularly interesting, since it implicitly implies what the constraint for the system is. \cite{yano2016size} provides a good approach to examine the assumptions of constraint (and therefore the value of $b$) using observational data. 

It should be noted that the generalized gamma distribution is derived when only one constraint of the power function of particle dimension is used. It is possible that more than one constraint exists or that the constraint functions $f_k(D)$ cannot be represented as power laws. Either way, the more general form of cloud PSD (Eq 26) can be used in such circumstances. The full potential of the MaxEnt for cloud physics applications will be realized after more understanding of the physical systems is gained. The development of idealized models to simulate the evolution of cloud particles can also provide another perspective, from which the application of MaxEnt may provide more theoretical basis on the appropriate constraint for the system that should be used.

%
\acknowledgments
The authors are supported by the office of Biological and Environmental Research (BER) of the U.S. Department of Energy Atmospheric Systems Research Program through grant No. DE-SC0016476 (through UCAR subcontract Z17-900029) and by the National Science Foundation (NSF) under grant AGS-1213311. The discussions with Hugh Morrison and Lulin Xue and the comments of three anonymous reviewers improved the quality of this paper considerably.

%
\appendix


\appendixtitle{List of variables and their definitions}

The variables used in this study are defined and summarized in Table A1. 

\begin{center}
\begin{table}
\appendcaption{A1}{List of symbols and their definitions.}
\centering
\begin{tabular}{ll}
\topline
\ Symbols & Definitions  \\
\midline
\ $\alpha$    & Prefactor of m-D relations  \\
\ $\beta$     & Power factor of m-D relations \\
\ $\gamma(s, x)$     & Lower incomplete gamma function \\
\ $\Gamma(x)$     & Gamma function \\
\ $\kappa$     & Scale factor between two length in two clouds \\
\ $\kappa_0$     & Lower limit of $\kappa_1$ \\
\ $\kappa_1$     & Scale factor between two volume in two clouds \\
\ $\bar \kappa$     & Scale factor between two coordinate system \\
\ $\lambda$         & The slope parameter in generalized gamma distribution in Eq (37) \\
\ $\lambda_1$     & The Lagrange multiplier for the first constraint \\
\ $\bar \lambda_1$     & The Lagrange multiplier relating to $\lambda_1$ by $\bar \lambda_1 = \alpha \lambda_1$ \\
\ $\tilde \lambda_1$     & The Lagrange multiplier realting to $\lambda_1$ by $\bar \lambda_1 = \frac{\alpha}{\beta} \lambda_1$ \\
\ $\lambda_k$     & The Lagrange multiplier for the $k$th constraint \\
\ $\mu$               & The shape parameter in generalized gamma distribution in Eq (37) \\
\ $\rho$               & Particle density \\
\ $\chi^2$               & The measure of goodness for a fit in Chi-square statistic \\
\ $a$                      & Prefactor of a general power law relations\\ 
\ $A$                     & The projected area of a cloud particle  \\
\ $b$                     & Power factor of a general power law relations in generalized gamma distribution in Eq (37) \\
\ $C$                    & The constant that relates to $\lambda_1$ through $C=C_0 exp(-\lambda_0)$ in Eq (10)  \\
\ $C_0$                 & The constant in Eq (10)  \\
\ $C_1$                & Constant in Eq (18) \\
\ $\bar C_1$          & Constant in Eq (19) \\
\ $\tilde C_1$         & Constant in Eq (22) \\
\ $D$                      & The maximum dimension of a cloud particle  \\
\ $e$                      & Euler's number, approximately equals 2.71828 \\
\ $E_i$                   & The $i$th kinetic energy state \\
\ $E$                      & Total kinetic energy of the particle system \\
\botline
\end{tabular}
\end{table}
\end{center}

\begin{center}
\begin{table}
\appendcaption{A1}{List of symbols and their definitions, Continued.}
\centering
\begin{tabular}{ll}
\topline
\ Symbols & Definitions  \\
\midline
\ $f_A(x)$               & The scaled invariant measure for $I_A(x)$\\
\ $f_B(x)$               & The scaled invariant measure for $I_B(x)$ \\
\ $f_k(x)$                & The $k$th constraint as a function of $x$  \\
\ $F_k$                  & The expected value of $f_k(x)$  \\
\ $IWC$                 & Ice water content \\
\ $I(x)$                   & The invariant measure \\
\ $I_A(x)$               & The invariant measure for cloud A \\
\ $I_B(x)$               & The invariant measure for cloud B \\
\ $I_R(x)$               & The invariant measure for coordinate system R \\
\ $I_S(x)$               & The invariant measure for coordinate system S \\
\ $k$                       & Constraint number \\
\ $L(x,\lambda_1, \lambda_2, ...,\lambda_n)$                  & Lagrangian function \\
\ $LWC$                & Liquid water content \\
\ $m$                     & The mass of a cloud particle  \\
\ $n$                     & Number of energy state in the ideal gas system  \\
\ $nc$                    & The number of constraints  \\
\ $nm$                   & The number of moments used for fitting  \\
\ $M_{obs,i}$          & The $i$th moment of the observed PSD  \\
\ $M_{fit,i}$            & The $i$th moment of the fit PSD  \\
\ $N$                      & Total number of ideal gas molecules  \\
\ $N_0$                  & Generalized gamma distribution parameter in Eq (37)  \\
\ $N(D)$                &  Number distribution function over size  \\
\ $N(m)$                &  Number distribution function over mass  \\
\ $N_i$                  & Total number of ideal gas molecules in energy state $E_i$  \\
\ $N_t$                  & Total number concentration  \\
\ $P_i$                  & Probability of ideal gas particles in energy state $E_i$  \\
\ $P(x)$               & Probability of $x$ state  \\
\botline
\end{tabular}
\end{table}
\end{center}

\begin{center}
\begin{table}
\appendcaption{A1}{List of symbols and their definitions, Continued.}
\centering
\begin{tabular}{ll}
\topline
\ Symbols & Definitions  \\
\midline
\ $S$                   & Gibbs/Shannon entropy  \\
\ $S_B$               & Boltzmann entropy  \\
\ $S_r$                & Relative entropy  \\
\ $TWC$            & Total water content \\
\ $v$                   & Cloud particle fall speed \\
\ $W$                  & The multiplicity representing the number of microscopic configurations  \\
\ $x$                   & A random state variable that describe the cloud particle  \\
\ $Z(\lambda_1, \lambda_2, ..., \lambda_n)$                   & Partition function  \\
\botline
\end{tabular}
\end{table}
\end{center}




%
%
%
\bibliographystyle{ametsoc2014}
\bibliography{references}

\end{document}